\documentclass[reprint,showpacs,preprintnumbers,amsmath,amssymb,superscriptaddress,bibnotes,prb]{revtex4-1}
\usepackage{graphicx}% Include figure files
\usepackage{dcolumn}% Align table columns on decimal point
\usepackage{bm}% bold math
\usepackage{color}
\usepackage{epstopdf}
\pdfoutput=1
\newcommand{\ham}{{\cal H}}

\begin{document}

\title{Spin nematics next to spin singlets}

\author{Yuto Yokoyama and Chisa Hotta}

\email{chisa@phys.c.u-tokyo.ac.jp}

\affiliation{Department of Basic Science, University of Tokyo, 3-8-1 Komaba, Meguro, Tokyo 153-8902, Japan}

\date{\today}% It is always \today, today,

\begin{abstract}
We provide a route to generate nematic order in a spin-1/2 system. 
Unlike the well-known magnon-binding mechanism, 
our spin nematics requires neigher the frustration effect nor a spin polarization in a high field or 
in the vicinity of a ferromagnet, but instead appears next to the spin singlet phase. 
We start from a state consisting of a quantum spin-1/2 singlet dimer placed on each site of a triangular lattice, 
and show that inter-dimer ring exchange interactions 
efficiently dope the SU(2) triplets that itinerate and interact, 
easily driving a stable singlet state to either Bose Einstein condensates or a triplet crystal, 
some hosting a spin nematic order. 
A variety of roles the ring exchange serves include the generation 
of a bilinear-biquadratic interaction between nearby triplets, which is responsible for the
emergent nematic order separated from the singlet phase by a first order transition.
\end{abstract}
\pacs{73.43.Nq, 75.10.Pq, 75.50.Ee, 75.40.Cx}

\maketitle

%*%*%*%*%*%*%*%*%*%*%*%*%*%*%*%*%*%*%*%*%*%*%*
{\em Introduction.---}
Challenges in modern magnetism have been to clarify the role of the intrinsic quantum effect 
in exotic phases of matter. 
Spin-1/2 liquids\cite{spinliq,spinliq2} that do not break any symmetry are characterized by 
the long range entanglement of their wave functions. 
Some phases break the symmetry quantum mechanically by forming the smallest entangled unit;  
a valence bond crystal is the long range order of a spin-1/2 singlet breaking translational symmetry\cite{majumdar}, 
and {\it spin nematics} is the SU(2)-symmetry-broken order 
of the quadrupole moment based on spin-1 pairs \cite{andreev84,papanicolaou}. 
The latter phase is our focus, and is established in a spin-1 bilinear-biquadratic Hamiltonian 
on square\cite{harada02,harada12,otimaa13,corboz17,corboz17-2} 
and triangular lattices\cite{tsunetsugu06,lauchli06,kaul12}; they appear whenever the biquadratic interaction 
$({\cal S}_i\cdot{\cal S}_j)^2$, overwhelms the bilinear (Heisenberg) term 
where ${\cal S}_i$ is the spin-1 operator. 
In fact, to entangle a pair of spin-1, one needs to exchange $({\cal S}^z_i,{\cal S}^z_j)=(+1,-1)$ with $(-1,+1)$ 
by the biquadratic term, such that the spin-1/2 singlets are formed by a Heisenberg exchange, flipping 
$(+1/2,-1/2)$ spins to $(-1/2,+1/2)$. 
Unfortunately, the spin-1 biquadratic interaction is usually much weaker than the bilinear term 
which makes the nematic phase as elusive as spin liquids. 
\par
Moreover, spin-1 is not the basic magnetic unit in condensed matter, 
since it appears only as a triplet pair of spin-1/2's of localized electrons. 
When the spin-1 is broken into pieces of 1/2, 
higher order exchanges among spin-1/2's are required to realize the spin nematics. 
The simplest one is the four-body ring exchange interaction. 
When this interaction is applied to the polarized spin-1/2 magnets 
in a strong magnetic field or near the ferromagnetically ordered phases, 
a nematic order indeed appears\cite{momoi12,nic06}, 
providing a good reference for a solid $\!^3$He\cite{fukuyama05,nema09}. 
There, the ring exchange was ascribed a role to enhance quantum fluctuation 
among competing magnetic orders\cite{nic06,lauhili05}. 
Besides, this interaction generates a gapped spin liquid in a spin-1/2 
triangular lattice antiferromagnet\cite{misguish98,misguish99,liming00}. 
The magnitude of ring exchange is enhanced near the Mott transition\cite{schmidt10}, 
which may explain the origin of the spin liquid triangular lattice Mott insulator\cite{morita02} 
possibly realized in the organic $\kappa$-ET$_2$Cu$_2$(CN)$_3$\cite{shimizu03}. 
The ring exchange further supports the anomalous thermal magnon Hall transport in 
a kagome ferromagnet\cite{katsura10}. 
%*%*%*%*%*%*%*%*%*%*%*%*%*%*%*
\begin{figure}[tbp]
\begin{center}
\includegraphics[width=7cm]{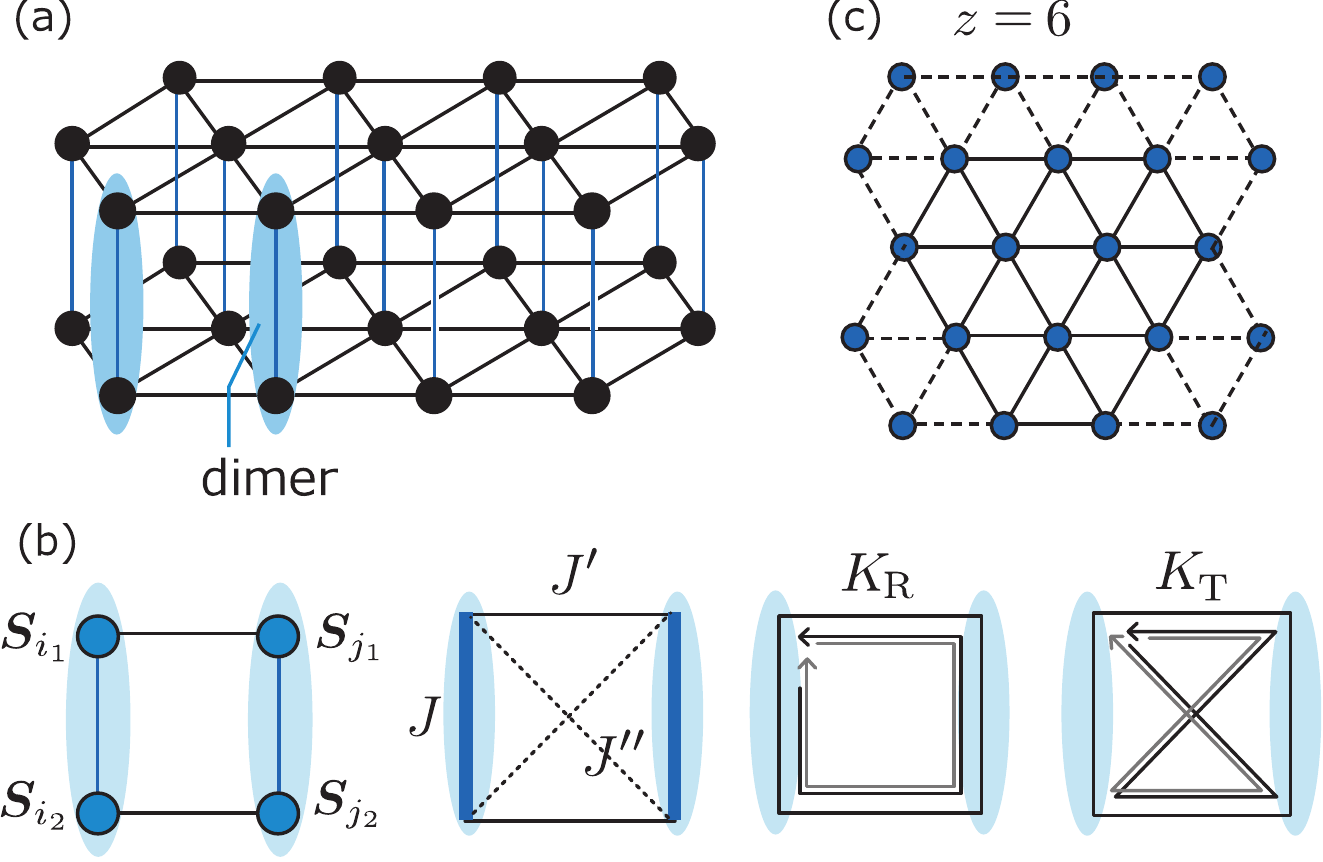}
\caption{(color online) (a) Triangular lattice based on $S=1/2$ dimers. Right panel is the top view of the lattice, 
and the $N=12$ sites used for the diagonalization. 
(b) Inter-site interactions: Antiferromagnetic Heisenberg exchange $J$, $J'$, and $J''$, 
four-body ring exchange ($K_{\rm R}$) and twisted ring exchange ($K_{\rm T}$). 
(c) Example of the operation of $K_{\rm R}$ and $K_{\rm T}$ to the ${\cal S}^z=\pm 1$ dimers. 
Spin-1/2's are permutated cyclically and become ${\cal S}^z=0$ and $\mp 1$ states 
by $K_{\rm R}$ and $K_{\rm T}$, respectively. 
The latter contribute to the  spin-1 biquadratic term. 
}
\label{f1}
\end{center}
\end{figure}
%*%*%*%*%*%*%*%*%*%*%*%*%*%*%*
%
%*%*%*%*%*%*%*%*%*%*%*%*%*
\begin{figure}[tbp]
\begin{center}
\includegraphics[width=6cm]{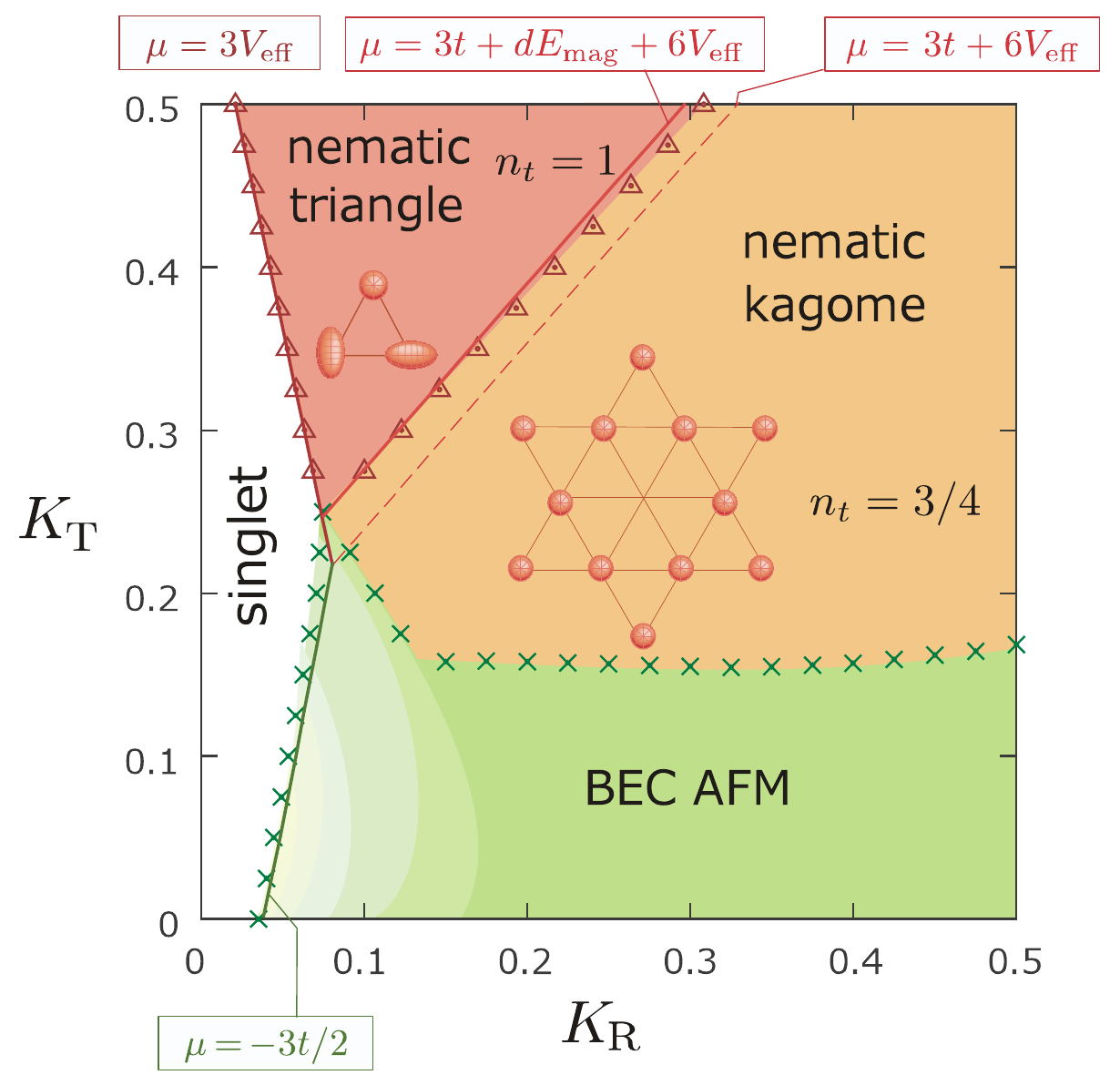}
\caption{(color online) Phase diagram of Eq.(\ref{ham0}) on the plane of $K_{\rm R}$ and $K_{\rm T}$, 
with $J=1$, $J''=0$ and $J'=2K_{\rm R}$.
The triangle and cross symbols indicate the first and second order transitions, respectively, obtained 
by analysis of exact diagonalization with $N=12$(24 spins). 
Solid lines are the analytical phase boundaries. 
By increasing $J'$ and $J''$, the nematic phase is stabilized toward 
smaller $K_{\rm R}$ and $K_{\rm T}$(see Supplementaly Material). }
\label{f2}
\end{center}
\end{figure}
%*%*%*%*%*%*%*%*%*%*%*%*%*%*%*
%
%*%*%*%*%*%*%*%*%*%*%*%*%*
\begin{figure}[tbp]
\begin{center}
\includegraphics[width=8.5cm]{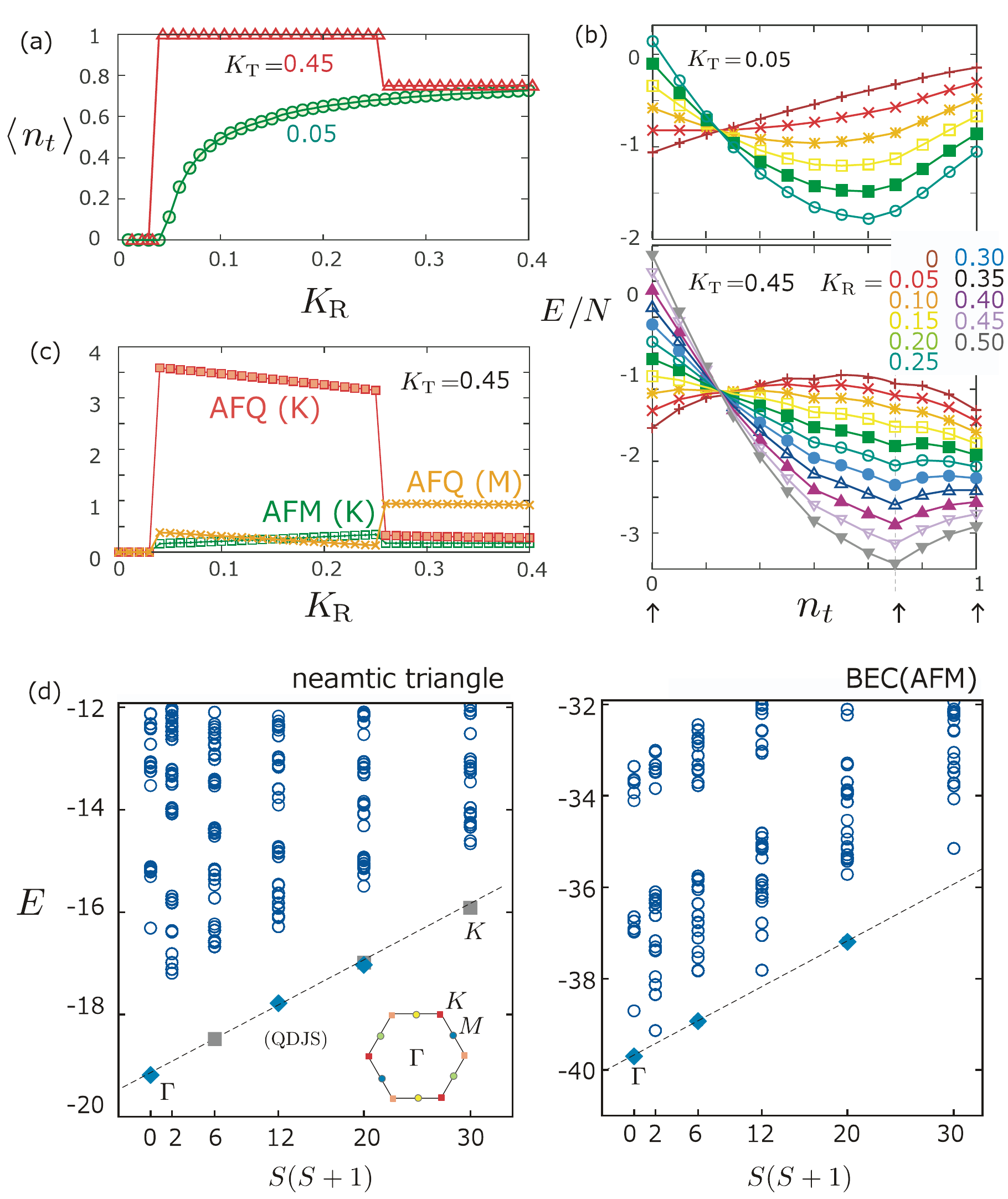}
\caption{(color online) 
(a) Density of triplets, $\langle n_t\rangle$ as a function of $K_{\bm R}$. 
In BEC, $\langle n_t\rangle$ changes gradually. 
A nematic triangle and kagome have exactly $\langle n_t\rangle=1$ and $3/4$, respectively. 
(b) Exact diagonalization energy per site $E/N$ of Eq.(\ref{ham0}) as a function of triplet density, $n_t$, 
at $K_{\rm T}=0.05, 0.45$ and $K_{\rm R}=0\sim 0.5$. 
(Upper panel) BEC phase with no anomaly. 
(Lower panel) Dips at $n_t=0,3/4,1$ indicate the three other stable phases in Fig.\ref{f2}, 
and the first order transitions between them. 
(c) Structural factor of $\langle Q_i\cdot Q_j\rangle$ at the $K$ and $M$-points on the Brillouin zone boundary, 
together with the ones for $\langle {\cal S}_i \cdot {\cal S}_j\rangle$, characterizing 120$^\circ$ magnetic ordering (AFM). 
(d) Tower of states in the nematic triangular ($K_{\rm R}=0.1, K_{\rm T}=0.45$)  
and BEC-AFM($K_{\rm R}=0.5, K_{\rm T}=0$)  phases. 
}
\label{f3}
\end{center}
\end{figure}
%*%*%*%*%*%*%*%*%*%*%*%*%*%*%*
Despite such rich physics relevant to the ring exchange, 
a clear-cut and systematic understanding of its role is still lacking. 
\par 
This Rapid Communication shows how the ring exchange serves to yield a variety of quantum phases. 
We consider spin-1/2 antiferromagnetically coupled dimers forming a triangular lattice. 
When the ring exchange interactions are transformed into a bosonic language, 
they simultaneously play different roles: chemical potential, hoppings, 
and repulsive/attractive interactions. 
Particularly, the one called {\it twisted ring exchange} flips the ${\cal S}_i$ pairs up side down, 
and contributes to the bilinear-biquadratic interactions. 
We obtain a rich phase diagram in the bulk limit that hosts 
Bose Einstein condensate(BEC) of $SU(2)$ bosons and 
nematic orders of triangular and kagome geometries. 
\par
{\em Model and Phase Diagram.---} 
Each lattice site consists of a pair of quantum spin-1/2 coupled by the antiferromagnetic Heisenberg interaction, $J$, 
as shown in Fig.\ref{f1}(a). 
When we consider only $J(=1)$, the ground state is an exact product state of singlets with energy $E_0=-NJ/4$. 
By introducing the inter-dimer interactions, 
our Hamiltonian is given as, 
\begin{align}
{\cal H} &= \sum_{i=1}^N J {\bm S}_{i_1}\cdot {\bm S}_{i_2}  +\sum_{\langle i,j\rangle,\gamma=1,2}\!
(J' {\bm S}_{i_\gamma}\cdot {\bm S}_{j_\gamma}
+J'' {\bm S}_{i_\gamma}\cdot {\bm S}_{j_{\bar\gamma}})
 \nonumber\\
& %%+\sum_{{\cal C}={\rm R,R',T}}
  +\sum_{{\cal C}} K_{\cal C} \Big(  \big(P_4+P_4^{-1}\big)
- \frac{9}{5}\sum_{(i_\gamma,j_{\gamma'})\in{\cal C}} {\bm S}_{i_\gamma} \cdot {\bm S}_{j_{\gamma'}} \Big)
\label{ham0}
\end{align}
where ${\bm S}_{i_1}, {\bm S}_{i_2}$ are the spin-1/2 operators forming the $i$-th dimer, 
$J'$ and $J''$ are the Heisenberg(bilinear) exchanges,  
and $K_{\cal C}$ denotes the four-body exchange (see Fig.\ref{f1}(b)). 
Along the two different closed loops ${\cal C}=$ R and T, 
the four spins permutate both clockwise ($P_4$) and anticlockwise ($P_4^{-1}$), 
which we call the ring exchange (R) and twisted ring exchange (T)\cite{supple}, e.g. as those shown in Fig.\ref{f1}(c). 
The last two-body term appears when deriving the four-body interactions by the perturbation from 
the Hubbard model at half-filling\cite{takahashi97,calzodo04,tanaka}. 
\par
Due to the ring exchange terms, a variety of phases emerges as shown in Fig.\ref{f2}; 
when $K_{\rm T}$ is small, $K_{\rm R}$ drives the system to the BEC of triplets. 
At larger $K_{\rm T}$, the nematic phases become dominant. 
Here, we stress that all the ${\cal S}_i^z=1,0,-1$ component of triplets equivalently join this BEC, 
which is thus {\it different from a magnon BEC} (carrying a net magnetic moment) 
that typically appears in spin singlet systems by the magnetic field\cite{giamarchi93,nikuni,ruegg03,rice02}. 
The spin nematics in a quantum spin-1/2 system known so far 
is based on a bound state of magnons created by the frustration effect, 
in a strong magnetic field or near the fully polarized ferromagnetic phase\cite{momoi12,chibukov91,hikihara08}. 
Our spin nematics {\it does not require such frustration}, and the overall feature of the phase diagram 
applies to square, honeycomb, and ladder systems as well\cite{yuto}. 
\par
{\em Bosonic representation.---} 
To understand the nature of the phase diagram, 
it is convenient to transform the basis in units of dimers rather than of spin-1/2's\cite{sachrev,picon08}. 
Each dimer hosts either a singlet (${\cal S}_i={\bm S}_{i_1}+{\bm S}_{i_2}=0$) 
or one of the triplets (${\cal S}_i=1$, ${\cal S}_i^z=1,0,-1$). 
Therefore, by regarding the singlet product state as a vacuum, %%$\cket{0}$,  
we introduce a bosonic operator, $b_{i,\alpha}^\dagger/b_{i,\alpha}$ with $\alpha=1,0,-1$, 
which creates/annihilates a triplet with ${\cal S}_i^z=1,0,-1$ on an $i$-th dimer. 
Equation (\ref{ham0}) is {\it exactly} transformed to 
$\ham = \ham_{tV} + \ham_{\rm mag} + \ham_{\rm pair}$ with 
\begin{eqnarray}
&&\ham_{tV}= 
\sum_{\langle i,j\rangle,\alpha}
\big( t(b_{i,\alpha}^\dagger b_{j,\alpha} + b_{j,\alpha}^\dagger b_{i,\alpha}) +V n_in_j  \big)- \sum_{i=1}^N  \mu n_i  \nonumber\\
&&\ham_{\rm mag}=\sum_{\langle i,j\rangle} 
\Big( {\cal J} ({\cal S}_i \cdot {\cal S}_j) + {\cal B} ({\cal S}_i\cdot {\cal S}_j)^2\Big)n_in_j \nonumber\\
&&\ham_{\rm pair}=\sum_{\langle i,j\rangle} P \big(b_{i,1}^\dagger b_{j,-1}^\dagger + b_{i,-1}^\dagger b_{j,1}^\dagger
- b_{i,0}^\dagger b_{j,0}^\dagger\big) +{\rm H.c.}, 
\label{htriplet}
\end{eqnarray}
where $n_i=0$ or 1 is the number operator of bosons under a hard core condition, and 
${\cal S}_i$ is the spin-1 operator on site-$i$ when it is occupied by a triplet, and fulfills 
${\cal S}_i^+=b_{i,1}^\dagger b_{i,0}+b_{i,0}^\dagger b_{i,-1}$, 
${\cal S}_i^-=b_{i,0}^\dagger b_{i,1}+b_{i,-1}^\dagger b_{i,0}$, 
and ${\cal S}_i^z=b_{i,1}^\dagger b_{i,1}-b_{i,-1}^\dagger b_{i,-1}$. 
The magnetic interactions, ${\cal J}$ and ${\cal B}$, will be discussed shortly. 
The parameters are given as, 
\begin{align}
& \mu=-J + \big(\frac{96}{5}K_{\rm R} - \frac{24}{5} K_{\rm T}\big),\\
& t=\frac{1}{2}(J'-J'')+K_{\rm R}, \\
& V= 4(K_{\rm R} -K_{\rm T} ), \\
& P=\frac{1}{2}(J'-J'')-K_{\rm R}. 
\end{align}
%*%*%*%*
In the present Rapid Communication, we take $J'=2K_{\rm R}$ and $J''=0$, in order to keep $P=0$, 
which allows for the exact evaluation of the phase boundaries, 
and the number of triplets is conserved. 
The role of $P\ne 0$ is mainly to enhance $\cal B$ and stabilize the nematic order 
(see Supplementaly Material), 
while the physics itself is not influenced qualitatively. 
\par
As the chemical potential, $\mu$, gets lower with $K_{\rm R}$, the bosons are doped to the vacuum (singlet). 
Meanwhile, there arises an interaction between doped bosons. 
Figure \ref{f3}(a) shows how the expectation value of the triplet density, $n_t\equiv \sum_i n_i /N$, develops, 
where $\langle n_t \rangle$ is the value that gives the minimum of total energy $E$ among all different $n_t$-sectors in Fig.\ref{f3}(b). 
At $K_{\rm R}>K_{\rm T}$, 
the kinetic energy gain due to $t$ favors doping the bosons, but the repulsive interaction $V>0$ does not, 
thus $\langle n_t \rangle$ increases gradually due to their competition. 
Contrarily at $K_{\rm R}<K_{\rm T}$, the attractive $V<0$ helps $\mu$ to dope triplets and there occurs 
a first order transition from the $\langle n_t\rangle =0$ to the $\langle n_t\rangle=1$ phase, 
which is also visible in the dip of energies at $n_t=0$ and 1 in Fig.\ref{f3}(b). 
Once all the sites are occupied by triplets, their spin-1's interact via $\ham_{\rm mag}$, 
which takes a well known form called the bilinear-biquadratic interaction, with 
${\cal B}= 2 K_{\rm T}$ and 
${\cal J}=(-K_{\rm R}+4 K_{\rm T})/5 +(J'+J'')/2$. 
In a triangular lattice, the spin-1 bilinear-biquadratic Hamiltonian 
is known to host a nematic long range order 
when ${\cal B}>{\cal J}$\cite{tsunetsugu06,lauchli06}, 
and the equivalent condition, $K_{\rm T}>3K_{\rm R}/2$, is actually fulfilled in our nematic triangular phase. 
\par
{\em Spin nematic order.---} 
The quadrupolar moment is the order parameter of the spin nematics 
and is described by a symmetric and traceless rank-2 tensor, 
$Q^{\alpha\beta}_j={\cal S}^\alpha_j {\cal S}^\beta_j+ {\cal S}^\beta_j {\cal S}^\alpha_j -2{\cal S}({\cal S}+1)/3 \delta_{\alpha\beta}$. 
We examined its two point correlation whose structural factor takes a peak at the $K$ and $M$-points at 
the Brillouin zone boundary, 
which are plotted as functions of $K_{\rm R}$ in Fig.\ref{f3}(c). 
A dominant peak at the $K$-point is consistent with the previously reported 
antiferro-quadrupolar ordering (AFQ) on the triangular lattice. 
The peak of the spin-spin correlation function at the $K$-point, 
characterizing the 120$^\circ$ antiferromagnetic (AFM) ordering is suppressed in these regions. 
%
%*%*%*%*%*%*%*%*%*%*%*%*%*%*%*
\begin{figure}[tbp]
\begin{center}
\includegraphics[width=8cm]{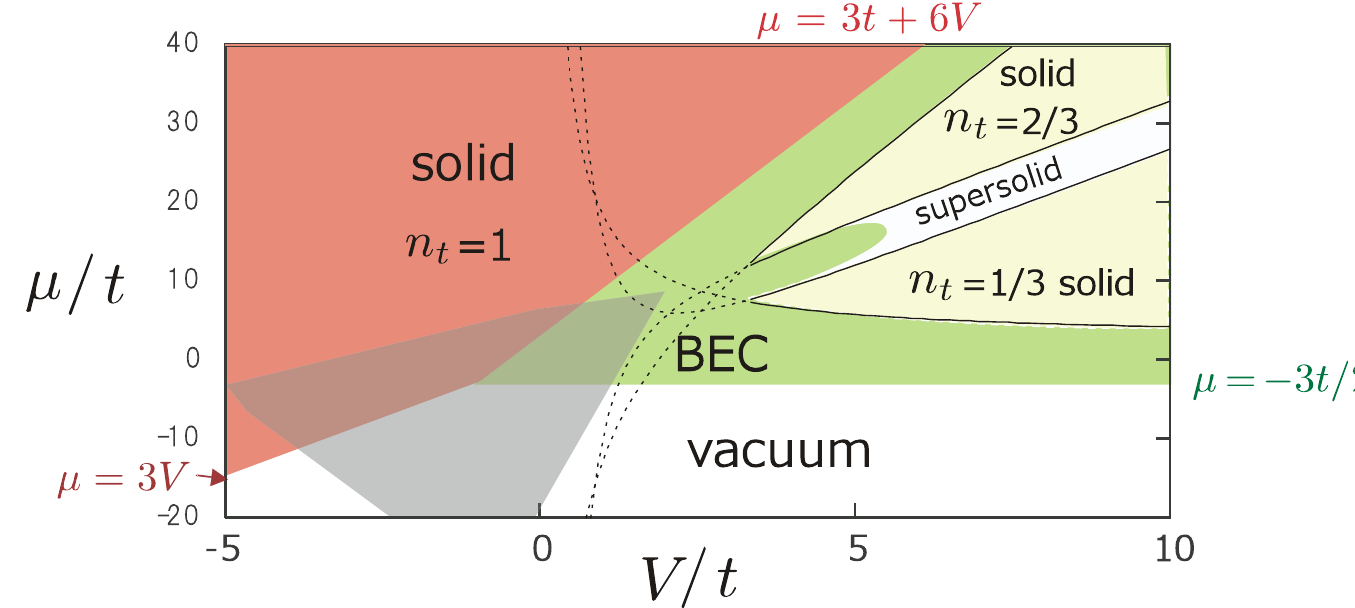}
\caption{(color online) Phase diagram of the hard core bosonic model $\ham_{tV}$ with $t>0$. 
The boundaries are evaluated analytically (for details see Ref.\cite{supple2}). 
The shaded area marks the parameter region realized in the phase diagram of Fig.\ref{f2}. 
}
\label{f4}
\end{center}
\end{figure}
%*%*%*%*%*%*%*%*%*%*%*%*%*%*%*
\par
To further confirm the existence of nematic long range order, 
we show the energy spectrum at $K_{\rm T}=0.5$ and $K_{\rm R}=0.1$ in Fig.~\ref{f3}(d). 
There actually appears a tower of low-lying energy levels well separated from 
the other excitations\cite{tower,bernu94}, 
and the symmetries of the quasidegenerate joint states (QDJS) belonging to different spin sectors 
follow those already known for the SU(2)-symmetry-broken spin nematics on a triangular lattice\cite{karlo}. 
We also give the same analysis on the BEC phase, finding that the antiferromagnetic long range order 
expected for $0<{\cal B}<{\cal J}$ in this region is realized as well. 
Notice that this phase is different from the AFM of a localized spin-1 system since the triplets are not 
fully occupied ($\langle n_t\rangle <1$). 
The slopes of the two towers are almost equal, further confirming the well known scaling behavior, 
$E_{\rm QDJS} \propto S(S+1)/N$. 
\par
{\em Twisted ring exchange.---} 
We need to understand why the biquadratic $\cal B$ in ${\cal H}_{\rm mag}$ (Eq.(\ref{htriplet}))
originates from the twisted ring exchange, $K_{\rm T}$, and not from $K_{\rm R}$. 
The spin-1 biquadratic term exchanges the up and down spin-1 pairs, 
changing ${\cal S}^z_i$ by $\pm 2$, 
while the Heisenberg (bilinear) term flips ${\cal S}^z_i$ 
only by the $\pm 1$.
Figure \ref{f1}(d) shows that when $K_{\rm R}$ is operated to the four spin-1/2's on a plaquette, they move cyclically, 
and transform the dimer spin $({\cal S}_i^z,{\cal S}_j^z)=(+1,-1)$ to $(0,0)$. 
Whereas, if the path is twisted, $K_{\rm T}$ can move the two spin 1/2's on one dimer to the other dimer at once, 
and flip the dimer spin $(+1,-1)$ to $(-1,+1)$, contributing to the biquadratic term. 
\par
The magnitude of $K_{\rm T}$ had been considered as small in a quantum spin system, 
as it originates from the fourth order perturbation in a Mott insulator\cite{takahashi97}. 
According to our evaluation\cite{tanaka}, the on-site Coulomb interaction $U$ against the inter-dimer transfer integral, 
$t_{ij}=t'$ or $t''$ 
should be $U/t_{ij} \lesssim 7$ in order to have $K_{\cal C}/J\gtrsim 0.1$, which is not too unrealistic. 
It is also shown that in the vicinity of the Mott transition, 
$U/t\sim 8$, the ring exchanges can be as large as $J/4$\cite{schmidt10}. 
We further mention that $P\ne 0$ works as an effective biquadratic term, 
thus a larger $J'-J''$ will stabilize the nematic phase than the one found in Fig.~\ref{f2}. 
(see Supplementaly Material). 
\par
{\em Instabilities.---} 
The phase boundaries of Fig.\ref{f2} can be determined half-analytically 
by examining the energetics of the hard core bosonic model, $\ham_{tV}$, in the bulk limit\cite{eggert11}. 
%For a moment, we forget about the spin-1 degrees of freedom. 
The phase diagram of $\ham_{tV}$ for $t>0$ is shown in Fig.\ref{f4}\cite{supple2}. 
Similar to the case of $t<0$ studied in the context of cold atoms\cite{troyer05,damele05,melko05,balents05}, 
the 1/3- and 2/3-filled crystal phases appear at large $V/t$, and the supersolid phases in between. 
The other parts are divided into the vacuum, bosonic BEC, and a solid, 
and their boundaries are {\it exactly} determined; 
The onset of the BEC from the vacuum is given by the kinetic energy gain of a single boson, $\mu=-3t/2$. 
The first order transition line between the vacuum and the solid takes place at $\mu=3V$, 
as the attractive interaction favors all the triplets to be doped at once by maximally 
gaining the energy $3V<0$. 
Finally, the instability of the solid against the BEC is evaluated by the energy of a doped hole, $-3t-6V$. 
\par
The shaded region in Fig.\ref{f4} covers the parameter range of Fig.\ref{f2}. 
The onset of BEC mapped to our model is, $111K_{\rm R}-24K_{\rm T}=5J$, 
in good agreement with the one from the exact diagonalization. 
In regions with higher boson densities, the bosons interact magnetically, 
thus we need to take account of the effect of $\ham_{\rm mag}$ terms on the energy of hard core bosons. 
For this purpose, we introduce an effective interaction including the corrections from the magnetic terms, 
$V_{\rm eff}=V + \langle {\cal J} ({\cal S}_i \cdot {\cal S}_j) + {\cal B} ({\cal S}_i\cdot {\cal S}_j)^2\rangle$, 
and evaluate its value by separately analyzing the spin-1 bilinear-biquadratic Heisenberg model. 
%% on the same triangular lattice. 
Originally, the upper left-half of the phase diagram, $K_{\rm T}>K_{\rm R}$ was the region with 
an attractive interaction $V<0$. 
However, this correction pushes the phase boundary upward, and the repulsive $V_{\rm eff}>0$ 
region starts just below the triangular nematic phase. 
The boundary between the singlet and the nematic triangle is given by $\mu= 3V_{\rm eff}$. 
\par
{\em Nematic kagome phase.---} 
There is another phase in the middle of the diagram with $\langle n_t\rangle=3/4$, 
characterized by the peak of the AFQ order at three $M$-points (see Fig.\ref{f3}(c),(d)). 
The spatial structure of this quadrupole order expected is a kagome geometry
that is realized by regularly depleting one quarter of the lattice sites of the triangular lattice. 
However, in a pure hard core bosonic model, there is no reason to favor such a kagome structure 
which is indeed absent in Fig.\ref{f4}. 
We thus reexamine the energy of the original Hamiltonian for all different $n_t$ sectors in Fig.~\ref{f3}(b);
there is a dip in the energy at $\langle n_t\rangle=3/4$, which competes with $\langle n_t\rangle=1$ 
when $K_{\rm T}\gtrsim 0.3$\cite{supple3}. 
This is in sharp contrast to the smooth $n_t$ dependence of $E$ in the BEC region. 
The competition between the three discrete fillings gives the first order transitions. 
The energy dip at $\langle n_t\rangle=3/4$ comes solely from the magnetic interaction, $\langle {\cal H}_{\rm mag}\rangle$, 
and not from $\langle \ham_{tV}\rangle$(Supplementaly Material Fig. S2), 
to be more precise, by the contribution from ${\cal B}({\cal S}_i\cdot {\cal S}_j)^2={\cal B}/2 (\langle Q_i \cdot Q_j\rangle -  \langle {\cal S}_i \cdot {\cal S}_j\rangle)+{\rm const.}$ 
This also suggests that the bosons remain BEC. 
We calculate the bond energy of the bilinear biquadratic Hamiltonian on a kagome lattice, 
and find that it is lower by $dE_{\rm mag}\sim 0.05$ compared to the same Hamiltonian on the triangular lattice. 
Thus, the phase boundary in Fig.~\ref{f2} is finally corrected to, $\mu=3t +dE_{\rm mag} + 6V_{\rm eff}$, 
showing excellent agreement with the ones obtained by the exact diagonalization. 
\par
{\em Remarks.---} 
The mechanism to generate a variety of phases in the spin-1/2 dimer model by the ring exchange interactions is fully fixed, 
by the exact transformation to a hard core bosonic language. 
We would like to stress the following points: 
The prototypes of emergent BEC's in spin systems were to dope magnons 
to a spin singlet state by the magnetic field,  
whereas, here, the ring exchange interaction serves as 
{\it a fictitious field that does not break the SU(2) symmetry}, 
and dopes the SU(2) bosons, not the magnons carrying magnetization. 
Once the bosons are doped, the bilinear-biquadratic interaction induced by a four-spin-exchange along the twisted path 
works to stabilize the nematic orders. 
The spin nematics basically requires an exchange of spin-1 bosons, thus was observed in the spin-1/2 system 
in a strong magnetic field (spin polarized state) or in the vicinity of frustrated magnetism, 
where the frustration played a key role to enhance the quantum fluctuation. 
Our spin singlet state is a trivial product state in the contour extreme limit of such a complication, 
thus one may feel it rather counterintuitive to have a nematic phase next to it. 
\par
The spin-1 hard core bosonic model can also 
be regarded as a strong coupling limit of the spinor boson systems studied in cold atoms. 
There, the bosons are softly exclusive on each site due to the on-site interaction, $U$, \cite{tsuchiya05,mila14}
and the second order perturbation from the $U/t\rightarrow \infty$ limit gives a biquadratic interaction 
between spin-1 bosons, and the nematic Mott insulating phase appears\cite{tsuchiya05,mila14,batrouni13}. 
A situation similar to man-made optical lattices is naturally realized in our quantum spin-1/2 model 
representing crystalline solids, 
as such a dimer system is actually quite ubiquitous in transition metals 
such as BaCuSi$_2$O$_6$\cite{sebastian}, Ba$_2$CoSi$_2$O$_6$Cl$_2$\cite{tanaka14}, and 
Ba$_3M$Ru$_2$O$_9$\cite{terra}. 
In Ba$_3M$Ru$_2$O$_9$, a nonmagnetic phase is actually found next to the singlet phase, 
and the relevance with our findings remains an issue to be clarified.

\begin{acknowledgments}
We thank Karlo Penc, Fr\'ed\'eric Mila, Shunji Tsuchiya, Ichiro Terasaki, and Katsuhiro Tanaka for discussions. 
This work is supported by  JSPS KAKENHI Grant Numbers (No. JP17K05533, No.JP17K05497, and No.JP17H02916).
\end{acknowledgments}

%[Reference]
%


\begin{thebibliography}
\bibitem{supple} There is another term $K_{\rm R'}$, comparable to $K_{\rm R}$ and $K_{\rm T}$ 
as shown in Ref.\cite{tanaka}, 
but plays an equivalent role with $K_{\rm R}$, thus is abbreviated in the present paper.
\end{thebibliography}
\end{document}